# Comparative Analysis of Cryptographic Key Management Systems

Ievgeniia Kuzminykh[1][0000-0001-6917-4234], Bogdan Ghita[2][0000-0002-1788-547X] and Stavros Shiaeles[3][0000-0003-3866-0672]

[1] King's College London, Strand, London, WC2R 2LS, UK
`ievgeniia.kuzminykh@kcl.ac.uk`
[2] University of Plymouth, Drake Circus, Plymouth, PL4 8AA UK
`bogdan.ghita@plymouth.ac.uk`
[3] University of Portsmouth, Portsmouth, PO1 3RR, UK
`stavros.shiaeles@port.ac.uk`

**Abstract.** Managing cryptographic keys can be a complex task for an enterprise and particularly difficult to scale when an increasing number of users and applications need to be managed. In order to address scalability issues, typical IT infrastructures employ key management systems that are able to handle a large number of encryption keys and associate them with the authorized requests. Given their necessity, recent years have witnessed a variety of key management systems, aligned with the features, quality, price and security needs of specific organisations. While the spectrum of such solutions is welcome and demonstrates the expanding nature of the market, it also makes it time consuming for IT managers to identify the appropriate system for their respective company needs. This paper provides a list of key management tools which include a minimum set of features, such as availability of secure database for managing keys, an authentication, authorization, and access control model for restricting and managing access to keys, effective logging of actions with keys, and the presence of an API for accessing functions directly from the application code. Five systems were comprehensively compared by evaluating the attributes related to complexity of the implementation, its popularity, linked vulnerabilities and technical performance in terms of response time and network usage. These were Pinterest Knox, Hashicorp Vault, Square Keywhiz, OpenStack Barbican, and Cyberark Conjur. Out of these five, Hachicorp Vault was determined to be the most suitable system for small businesses.

## 1 Introduction

The complexity and reliability of any cryptosystem is based on the use of cryptographic keys. The key exchange is one of the mechanisms at core of the process and it ensure confidentiality when exchanging information between users and its behaviour is well-explained for small systems. However, in larger IT infrastructures, reaching hundreds or thousands of users, the process of handling the cryptographic keys for individual business applications is an increasingly difficult task and poses significant challenges as manual decentralized control is expensive and error prone. Such complex environments, including a large number of systems, group accounts, and users associated with them require a convenient and effective way to manage them. The solution is to use a cryptographic key management system (KMS) that provides a unified interface for managing keys, increase security of the enterprise network, provide scalability, and minimize human errors [1]. If an IT infrastructure does not include a sufficiently reliable management of key information then, having taken possession of it, an attacker may also gain access to the stored information, user accounts, their associated information and any databases [2].

A cryptographic KMS is a centralized system that provides key generation, key storage and key distribution, as well as automatic expiration, updating, re-placement, backup and revocation of keys, all for a wide range of applications [3, 4]. A typical example of a large scale KMS is a public-key infrastructure PKI, which uses hierarchical digital certificates for authentication and public keys for encryption.

Given their necessity for current IT infrastructure, several key management methods and tools are available on the market, varying in terms of cost, complexity and use cases. In this context, selecting the most suitable option becomes a challenge, since some of these solutions may be impractical or difficult to implement; as a result, the process of selection can be time consuming or unfeasible for an organisation that does not have enough resources to conduct such an assessment. This study aims to address this problem by comprehensively comparing the existing key management systems that match a set of attributes important for small businesses, such as simplicity of installation, ease of usage, performance and price, that may take precedence over scalability and security.

## 2   Key Management System Architecture and Features

A Key Management System is used to centrally distribute and store all keys used by an organisation and may take various forms, ranging from free small applications that run on conventional computer equipment to complex hardware solutions. Simple, open source solutions often rely on a regular database server for storage that stores keys encrypted in the database. However, due to the importance of the key management system, an appropriately designed system should include a hardware security module for key processing, or at least consider such an option [5].

The typical components of a KMS include the KMS Server, the KMS Client, the Hardware Security Module (HSM) and a database [6, 7], as summarised in the architecture on Fig.1.

The *KMS Server* is the central part of the system, where the actual key management takes place; this module is also responsible for the all operation related to the encryption key lifecycle, from generation, then activation, expiration and then destruction, as well as the key allocation against targets [8]. In order to deliver its functionality, the KMS Server connects to its dedicated HSM and a database to provide key management and key requesting services.

The *KMS Client* provides a graphical user interface for users to manage and operate the KMS server. While less critical in terms of process and security provision, this entity is essential for the customer, because it reflects system usability, attractiveness, and simplicity.

The *Hardware Security Module* is used by KMS to ensure the quality of keys generated and the protection of these keys while in storage or in transit. HSM can perform a number of important security-related cryptographic operations such as encryption, hashing, digital signing, and Message Authentication Codes (MAC). Additionally, sophisticated techniques can be used to ensure that keys are never present in unencrypted form in server memory or client machines [5]. Typically, an HSM is installed inside a server or within an Ethernet cluster within corporate network.

The *Database* stores all data, but sensitive information such as keys and key components are also encrypted under the master key generated by the HSM. Keys only exist in clear text inside the HSM. Other sensitive data, for example settings and logs, are integrity protected by a hardware MAC key so that the data cannot be edited without the server. Different solutions may choose to implement their respective databases using any preferred database management systems, such as Oracle, MySQL, PostgreSQL, DB2, Redis, etc.

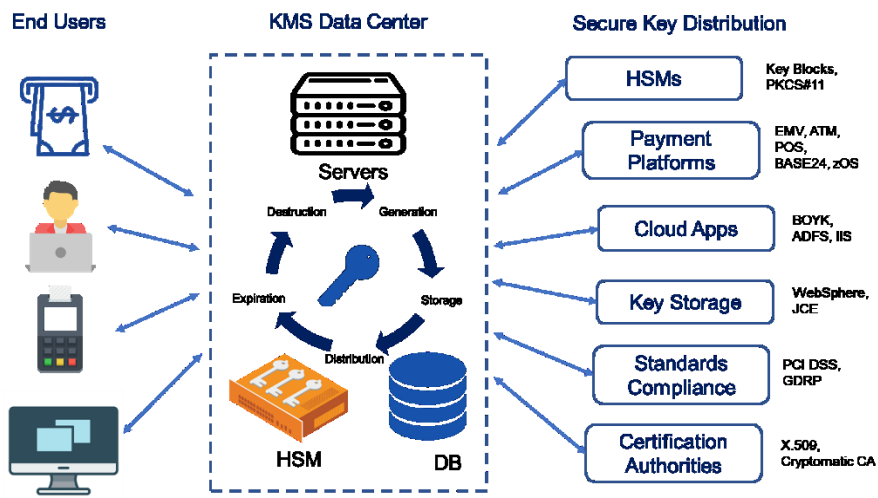

**Fig. 1.** Architecture of KMS with functions and applications.

The variety of use case where a KMS can be deployed and where keys are necessary to be securely handled makes it likely that customers have many reasons to use KMS in their business-critical processes. Some typical use cases include payment platforms, such as ATM or POS systems with Remote Key Loading (RKL) [9], Europay, Master-Card and Visa (EMV) keys for card issuance and authorization [10], cloud applications such as Bring Your Own Key (BYOK) to cloud environments [11, 12], HSM application keys, e.g. Atalla, Thales, etc., keys for data protection [13], e.g. PCI DSS, GDPR compliance, and Identification and Authentication Management Systems (IAMS) [14,15], or X.509 certificates for PKI, web servers, IoT devices [16-18].

## 3 Research Methods

The previous section provided an overview of the underlying concepts of KMS. Aside from highlighting a number of necessary characteristics, such as the presence of the four main entities, it also raised a number of issues that do not war-rant an increased level of security but provide other desired or supporting functionality, such as protection of the KMS infrastructure itself or improved usability and user experience.

Based on these concepts, this section proposes a three-stage process to identify, reduce, and comparatively analyse the existing KMS solutions. The identification stage will involve an extensive search to highlight existing solutions. In order to reduce the analysis task, the second stage will use a consistent set of the security attributes to investigate their impact for a typical organisation to shortlist and quantitatively assign their importance for a unified score. Finally, the third stage will involve a combination of feature analysis and benchmarking of the shortlisted solutions.

### 3.1 Literature Review

For literature review, various search queries were used to look for scientific articles and technical documentation regarding key management systems that currently available to handle software secrets. For searching both scientific and non-scientific search engines were used. The keywords used to derive start set were: "encryption key management system", "cryptographic key management system", "software key management system", "enterprise key management system". Among the scientific papers, only a few publications describing the KMS were found, the main part of the search results was either on the websites of the manufacturers of this software KMS tool, or on specialized source code sharing platform with software, such as Github.

The following inclusion and exclusion criteria were used to identify a set of KMS for screening the results of the search:
- IC1: The KMS solution should be software based.
- IC2: The KMS solution should be currently available for download, subscription or purchasing.
- IC3: There should exist description of the system, minimal documentation/specification.
- EC1: The project (e.g. on the GitHub) is apparently dead (i.e., no recent comments, updates or support contact information).
- EC2: the KMS tool is not presented in English.

### 3.2 Selection Process

As previously mentioned, a large number of existing KMS products have been proposed and are currently available as implemented solutions for organisations to adopt. The search proposed in the previous section would therefore yield a large number of results, all matching the context of a small organisation, but without considering the actual abilities of the respective products. In order to prime the process of evaluating them, a number of discriminating criteria are proposed in this section to derive the systems that are likely to best align to the requirements of typical small and medium organisations; these criteria are summarised below:

1. Cost. While technical capabilities are critical, small organisations are likely to face a more pragmatic challenge in terms of affordability. This criterion may exclude indeed more mature, complex solutions, but it will also represent the prime selection choice for smaller organisation.
2. REST API. A KMS must be integrated with the existing IT infrastructure, interface, and messaging, more specifically with the existing applications and users that must be connected. Having a REST API does not guarantee a smooth interconnectivity, but without it most organisations will perceive the integration task as a rather daunting one and may decide to opt out.
3. Perform an authentication of users and inside itself (who logged in and what actions are performed). All access to systems is required to be authenticated, preferably using client certificates.
4. Access control policy (how, when and by whom a key is accessed). Beyond key handling, a KMS must have the ability of authorization the subject of access using an access control list.
5. Logging. A complete solution should not only successfully manage keys but also be able to review and audit them against the authentication and access control list.
6. Secure communication. This is an expected criterion, as it links directly with the strict requirement for encryption, preferably suing a prior, approved standard.
7. Local storage. This is unlikely to provide substantive performance improvements, but it will reduce the dependency for external resources.
8. Backup. Given keys underpin the entire infrastructure, it is vital to have in place backup mechanisms that can ensure the survivability of the keys as part of the disaster management policy.
9. Scalability. The scalability challenges closely correlate with the size and complexity of business; one typical example is the computational requirements while performing cryptographic operations as the number of keys, users and applications increase.

The aim of these criteria is to reduce the number of solutions to a manageable figure and follow with a comparative analysis only on these systems.

### 3.3 Comparative analysis

The KMS tools selected in the previous step were analysed using a set of attributes. Each attribute corresponded to a question with an assigned score, ranging from 1 point for low-impact, to 2 points for relevant, and 3 points for critical attributes. Given the paper focuses on the small and medium business needs, financial resources were highlighted as a priority and, hence, the analysis was restricted only to free-of-charge solutions. Similarly, the complexity of the KMS tool installation and managing process was also perceived as a significant issue and hence associated with a significant impact.

To measure installation and managing complexity, performance, and usability, each key management system was integrated in a C# test application which represents unit tests.

The complexity of the implementation was defined as the time, steps and knowledge necessary for the complete implementation of the solution, and whether additional actions such as changing or adapting the initial source code are required. Performance was measured based on the response time — the time it takes for retrieving a key from the KMS's storage — and network usage that shows amount of data that is sent over the network when retrieving a key. Ease of use implies qualitative indicators of how easy it is to use the system after its implementation, to monitor, read the logs, how convenient the user interface is. In order to evaluate usability, it is necessary to conduct a survey with physical respondents who used or currently are using the encryption key management tools and who can help to estimate an usability, as well as pros and cons of the solution. However, conducting a survey on each tool requires more time for research and, thus, was deselected.

The list of the attributes is presented in Table 1.

Table 1. List of attributes for comparative analysis of KMS.

| 3 points | 2 points | 1 point |
| --- | --- | --- |
| Secure storage | Multiple authentication methods | Is the KMS beneficial of having unit tests in the source code? |
| Audit logs | Automated start | Popularity in the developer community (based on the number of stars and watches) |
| Access control | Comprehensive documentation | |
| No known vulnerabilities | HSM support | |
| High impact in the community (low ratio of open/total issues) | Available for commercial use | |
| Actively maintained and developed? (based on the number of recent commits) | Technical support for developers | |

Each KMS can be evaluated by adding the scores from each attribute. If the answer yes to the question then the points are added to the score, otherwise nothing is added. In relation to the ratio of open/total issues, the points are added when the ratio is below average, calculated among all the compared KMS. Similarly, the popularity points are awarded for systems that have a number of commits higher than the average. Performance results are evaluated with high score of 3 points each in case when values are less than average for response time and network usage.

### 3.4 Experimental environment

In order to evaluate the parameters of the installation speed and performance of key management tool, each of the five KMSs was installed on a virtual machine Ubuntu 18.04 that had characteristics with one CPU, 4096 GB RAM and 60 GB operative memory. The physical machine for running VMs had hardware characteristics with i7-8550U CPU and 16 GB RAM. Key management tools were configured, and had a way to interact with test applications via REST API. The REST calls were initiated by the console test applications in order to perform different operations as requesting a key, storing a key, retrieving a key, authentication of the communication with the specific KMS. The execution performance was measured during retrieving a key operation, and was marked using BenchmarkDotNet tool [19]. Another performance metric, the network usage, was measured using Process Monitor, as well as all network traffic that was sent between the key management system and test application.

## 4 Results and Analysis

The comprehensive search based on the IC and EC criteria from Section 3.1 produced a list of 58 software based key managements tools, of which 32 were open source code and 26 were closed source.

Further assessment of key management tools was made on matching important requirements for generating, storing, providing access and transmission of secrets, as described in Section 3.2. An analysis of these requirements provides an idea of how well key management systems may fit the needs of small and medium-sized businesses. Table 2 and Table 3 show a general comparison of the functions of all identified KMSs with respect to the SME requirements.

Table 2. Open source KMSs evaluation. Empty cells mean Yes.

|  | Free of charge | REST API | Auth with certificate | Access control policy | Audit logs | Secure communication | Keys on-premises |
|---|---|---|---|---|---|---|---|
| Ansible Vault |  | No | No | No | No | N/A |  |
| Bastillion-io Bastillion |  | No | No |  |  |  |  |
| Chef Vault | No | No | No |  | No | N/A |  |
| Cloudflare Red October |  |  | No |  | No |  |  |
| Codahale Sneaker | No | No | No |  |  |  | No |
| Cyberark Conjur |  |  | No |  |  |  |  |
| Docker Biscuit | No | No | No |  | N/A | N/A |  |
| Docker Secrets |  | No | No | No |  |  |  |
| EnvKey | No |  | No |  |  |  | No |
| Flix- Keeto |  | No | No |  |  | N/A |  |
| FreeIPA |  | No |  |  | No |  |  |
| Fugue CredStash | No | No | No |  |  | N/A | No |
| GnuPG |  | No | No |  |  |  |  |
| Hashicorp Vault |  |  |  |  |  |  |  |
| LatFchset Custodia |  |  | No | No |  |  |  |
| Lyft Confidant |  | No | No |  |  |  |  |
| Manifold Torus |  | No | No |  | No |  | No |
| Meltwater Secretary |  | No | No |  |  |  |  |
| Mozilla SOPS | No | No | No |  |  | N/A |  |
| Neat S.r.l. Kmc-Subset137 |  | No | N/A | N/A | N/A | N/A | N/A |
| Oleiade Trousseau |  | No | No | No | No | N/A |  |
| OpenSSH |  |  |  |  | No |  |  |
| OpenStack Barbican |  |  |  |  |  |  |  |
| Pinterest Knox |  | No |  |  |  |  |  |
| Poise Citadel | No |  | No |  | No |  | No |
| PrivacyIDEA |  | No | No |  |  |  |  |
| Schibsted Strongbox | No | No | No |  |  |  | No |
| Shopify EJSON |  | No | No |  | No | N/A |  |
| Shyiko Kubesec |  | No | No |  |  |  |  |
| Square Keywhiz |  |  |  |  |  |  |  |
| T-Mobile T-Vault |  |  | No |  |  |  |  |
| XOR Data Exchange Crypt |  | No |  |  | No |  |  |

**Table 3.** Evaluation of closed source KMSs regarding to the requirements of small businesses. Empty cells mean Yes.

| | Free of charge | REST API | Auth with certificate | Access control policy | Audit logs | Secure communication | Keys on-premises |
|---|---|---|---|---|---|---|---|
| Amazon AWS KMS | No | | | | | | No |
| Amazon AWS Secrets Manager | No | | | | | | No |
| AppViewX CERT+ | No | | N/A | | | | N/A |
| Bloombase KeyCastle | No | | N/A | | | | |
| Chef Vault | No | No | No | | No | N/A | |
| CipherCloud Key Management | No | | N/A | | N/A | | N/A |
| Cryptomathic Crypto Key Management System | No | | N/A | | | | N/A |
| Egnyte | No | | | | | N/A | |
| Fornetix Key Orchestration | No | | N/A | | | | N/A |
| Futurex Key Management Servers | No | | N/A | N/A | N/A | N/A | |
| Gemalto Safenet Virtual KeySecure | No | | No | | | | |
| Google Cloud KMS | No | | | | | | No |
| Hancom SKM | No | | N/A | | | | |
| Hytrust KeyControl | No | | N/A | | | | |
| IBM Security Key Lifecycle Manager | No | | N/A | | N/A | | N/A |
| KeyNexus Key Management as a Service | No | | N/A | | | | |
| Kryptus KNET | No | | N/A | | N/A | | |
| Microsoft Azure Key Vault | No | No | | | | | No |
| Oracle Key Manager | No | No | N/A | | N/A | | |
| Oracle Key Vault | No | No | N/A | | | | |
| Quintessence qCrypt | No | | N/A | | | | |
| SSH.com Universal SSH KeyManager | No | No | N/A | | | | |
| Thales Vormetric Data Security Manager | No | | N/A | | | | |
| TokenEx | No | | | | No | N/A | |
| Townsend Security Centralized Encryption KMS | No | | N/A | | | | |
| Unbound KeyControl | No | | N/A | | | | |
| Zettaset Xcrypt | No | | N/A | | N/A | N/A | |

The information in Table 2 and 3 allowed selecting the systems that satisfy all the mandatory requirements: be open source and free of charge, include a REST API for user communication and other tools for analysis and visualization of acquired data in real time, support user access control, be able to record audit logs that allow a timely responses when an incident takes place, provide secure communication for key exchange, store keys locally cached on the client ma-chine to prevent outages if server side is unavailable. The authentication through certificates is considered preferable but not a requirement.

The analysis of key management systems summarised by Table 2 determined that only six tools correspond to the requirements and are suitable for further more comprehensive analysis: Cyberark Conjur, Hashicorp Vault, OpenStack Barbican, Pinterest Knox, Square Keywhiz, and T-Mobile T-Vault. Given that the T-Mobile T-Vault is an extension of the Hashicorp Vault, it was excluded from the comparison.

Following the shortlisting, the selected five key management systems were analysed for a set of attributes and for the complexity of implementation, discussed in Section 3.3. The analysis was based on data from the research studies [20, 21], Github repositories and technical documentation of each tool [22-27]. The comparison results are summarised in Table 4, where the attributes are arranged in order of their weight.

**Table 4.** Comparison of the KMS. Empty cells mean Yes. The data was collected 2020-06-07.

|  | Conjur | Vault | Barbican | Knox | Keywhiz |
|---|---|---|---|---|---|
| Secure storage (3p) |  |  |  |  |  |
| Audit log (3p) |  |  | No |  | No |
| Access control (3p) |  |  |  |  |  |
| No known vulnerabilities (3p) |  |  | No | No | No |
| Open/Total issues (3p) | 27 | 14.1 | 17.3 | 0 | 27.7 |
| Recent commits (3p) | 74 | 158 | 16 | 6 | 39 |
| Response time (ms/op.) (3p) | 6.95 | 1.56 | 83.37 | 0.78 | 2.31 |
| Network usage (kB/op.) (3p) | 3.63 | 3.93 | 1.76 | 3.65 | 2.86 |
| Multiple auth methods (2p) | No |  |  | No |  |
| Start automatically (2p) |  | No |  |  |  |
| Well-written documentation (2p) |  |  |  | No | No |
| HSM support (2p) |  |  |  | No | No |
| Open for commercial use (2p) |  |  |  |  |  |
| Technical support (2p) |  |  |  | No | No |
| Unit tests (1p) |  |  |  |  |  |
| Popularity (1p) | 422 | 165552 | 226 | 870 | 2294 |
|  |  |  |  |  |  |
| Total score | 29 | 33 | 25 | 23 | 20 |

The analysis shows that all KMSs can be easily integrated in the existing infrastructure through an API client, perform access control and logging information about client actions and secrets, as stated during the selection criteria for the full list of KMSs. There is a small variation, as Keywhiz and Barbican do not log information regarding the client identity who access the resource or request the secret. In addition, all of the KMSs could be used for commercial purposes, allow to test source code with unit tests and support multiple secure storage backends, as databases (MySQL, PostgreSQL, etc), cloud storages and file system.

From a safety perspective, Knox, Barbican, and Keywhiz have no published security vulnerabilities, while Vault and Conjur had few but they are fixed and patched already in newer versions. Knox currently has no open issues, and the ratio of open issues to total issues for Vault and Barbican is below the mean; Conjur and Keywhiz have more open issues than the average amongst the other KMS. The amount of commits made over the most recent month was below the mean ratio for Knox, Keywhiz and Barbican and above the average for Vault and Conjur.

Moving onto performance, the analysis showed that Barbican has a very high response time during when extracting the secrets from the storage, one order of magnitude higher than the other analysed KMSs. Vault, Knox and Keywhiz showed quite well response time values of few seconds and Conjur had a slightly higher response time but still below average. In terms of the generated traffic, the network usage analysis showed that Knox, Conjur and Vault key management tools send more data over the network than the average.

In terms of their features, there was some variation among the set. Knox, Vault support authentication using multiple different types of methods. Knox supports three auth methods: Mutual TLS, Github Access Tokens and SPIFFE but, during the setup process, only the Github Access is available. Vault supports Github credentials, tokens, certificates and credentials to cloud providers. Square Keywhiz supports authentication with password and certificates. Openstack Barbican supports using SAML, user credentials, tokens and certificates. Conjur can authenticate using account ID and API key, LDAP and AWS credentials. All compared KMSs, except Vault, are able to start automatically using a command at server boot up. Vault requires unlocking of the server using a set of keys. Conjur, Vault and Barbican support HSM as a storage backend, and Knox and Keywhiz do not support HSM.

Almost all KMSs have well-written documentation, with thorough explanations of the features, configurations and various functionalities. However, Knox lacks the technical details on how to use the provided API, and Keywhiz only explains how to use the tool in the development mode but lack guidelines for making it production ready. It has also broken external links. Barbican lacks a logical structure that makes it difficult to find a right section. Installation complexity analysis showed that during installation the Knox source code requires a lot of editing to finally operate correctly. Vault, Keywhiz, Barbican and Conjur are relatively easy to install, through configuration files and running commands. Pinterest Knox and Keywhiz do not provide any technical support while Vault and Barbican offer it through the open IRC channel. Finally, from the perspective of popularity, Barbican, Knox and Conjur scored lower than Vault and Keywhiz, which are significantly higher in terms of ranking amongst similar applications.

Altogether, the analysis showed that Hashicorp Vault has the highest final score, while Square Keywhiz got the lowest score among key management tools.

## 5 Discussion and Conclusions

Based on an extensive search, a comprehensive list of key management systems was identified as viable KMS alternatives for small business. Totally, out of 58 tools, 32 were open source and 26 were closed source. Open source KMSs varied in terms of complexity: from tools for in-house usage, like Pinterest Knox, to KMSs that are part of company business plans, like Hachicorp Vault. A similar variation of complexity can also be seen in the closed sourced KMSs, with some are available as a service, such as

Amazon AWS KMS or Microsoft Azure Key Vault, and others being hosted locally, such as Thales Vormetric Data Security Manager.

Each organisation does have a specific set of KMS requirements, but in the paper we proposed a common set likely to be specific for all SMEs. First of all, the chosen level of logging favours traceability over performance and disk usage, to make sure that all actions can be traced back to a user. Secondly, the communication between clients and the KMS needs to be secure to prevent leakage due to snooping. Next, for usability, the preference is for the KMS were to start automatically on boot, since such a KMS would have a less negative impact on its clients during an unplanned reboot. Finally, the security policy also states that the keys and secrets should be on the premises because utilising key management as a service from third-parties requires trust in the service provider; this can bring usability advantages but may weaken the confidence.

For more detailed analysis of selected five key management tools the qualitative attributes were defined. Among them there were the quality of installation guides, popularity of tool among community, implementation and usage simplicity, availability of technical support, and the variety of authentication methods.

The evaluation concluded that Hachicorp Vault scored highest as it is equipped with well-written documentation and educational guides, and due to the wide range of authentication methods and storage backends that it includes, which makes it suitable for multiple different businesses. For best practice, Hachicorp Vault should be configured and installed with its own storage backend, named Consul, on physical machines with restricted access. Further, Hachicorp Vault should only communicate securely using TLS; the machines, Vault, and other running services should be actively updated to mitigate unauthorized access. Compared to the other KMSs, Vault sends more data than average during key exchange because it includes more metadata.

One major reason for the success of Vault was its business model. Hachicorp offers Vault based on a freemium business model, where the entry model of the version is free and additional tiers costs. This means that Vault can be extended in the paid version, for example, should HSM support be required.

The results of the paper can be used by small businesses as a guideline of how to perform an evaluation themselves. The results can also be used either partly or fully by a small company if their requirements of a KMS are partly or completely equal to the requirements outlined in this study.

As part of future work, commercial, closed-sourced KMSs should be investigated, in order to get a better understanding of the differences (if any) between paid and free of charge KMS, and thus deciding if they justify the cost. Another way of extended study is evaluating more systems that give a broader perspective of the use cases and limitations from various KMSs. It would also provide a better comparison for selected attributes. Moreover, the attributes applied could also be extended and include post installation comparison as how easy a system can be updated, backed up and restored.

## Acknowledgement

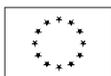 This project has received funding from the European Union Horizon 2020 research and innovation programme under grant agreement no. 786698 and no. 833673. This work reflects authors view and Agency is not responsible for any use that may be made of the information it contains.

# References

1. 2020 Global Encryption Trends Study. Ponemon Institute Research Report (2020)
2. Sinha, V. S., et al.: Detecting and mitigating secret-key leaks in source code repositories. In: 12th Working Conference on Mining Software Repositories (MSR), pp. 396–400. IEEE/ACM, Florence (2015)
3. Björkqvist, M. et al.: Design and Implementation of a Key-Lifecycle Management System. In: Sion R. (eds) Financial Cryptography and Data Security. FC 2010. Lecture Notes in Computer Science, vol 6052. Springer, Berlin, Heidelberg (2010)
4. Selecting the right key management system. Cryptomathic White Paper (2019)
5. Attridge, J.: An Overview of Hardware Security Modules. SANS Institute. Information Security Reading Room (2002)
6. Biggs, A., Cooley, S.: Management Service Architecture, IETF Internet draft (2015)
7. Mogull, R: Understanding and Selecting a Key Management Solution. Securosis LLC (2013)
8. Allen, C.: Exploring the Lifecycle of a Cryptographic Key (2018). https://www.cryptomathic.com/news-events/blog/exploring-the-lifecycle-of-a-cryptographic-key-. Last accessed 17 June 2020
9. Cryptera. Understanding Remote Key Loading, https://www.cryptera.com/wp-content/uploads/2014/07/Cryptera_WP_Understanding-RKL_To-Launch.pdf. Last accessed 10 June 2020
10. EMV Key Management. Cryptomathic White Paper (2017)
11. Kumar, V., Sharma, I.: Bring-your-own-encryption: How far are we? In: 11th International Conference on Industrial and Information Systems (ICIIS), pp. 672-677, Roorkee, (2016)
12. AlBelooshi, B., Damiani, E., Salah, K., and Martin, T.: Securing cryptographic keys in the cloud: A survey. IEEE Cloud Computing **3**(4), 42–56 (2016)
13. Mogull, R: Pragmatic Key Management for Data Encryption. Securosis, LLC (2012)
14. Kuzminykh, I., Fliustikova, M.: Mechanisms of ensuring security in Keystone service. Problems of Telecommunication **2**(25), 78-96 (2019)
15. Sitaram, D., Harwalkar, S., Simha, U., Iyer, S., and Jha, S.: Standards based integration of advanced key management capabilities with openstack. In: IEEE International Conference on Cloud Computing in Emerging Markets (CCEM), pp. 98–103. IEEE, Bangalore (2015)
16. White, C., Edwards, S.: Server-client PKI for applied key management system and process. US Patent US10560440B2 (2020)
17. Kuzminykh, I., Yevdokymenko, M., Sokolov, V.: Encryption Algorithms in IoT: Security vs Lifetime. Data-Centric Business and Applications. LNDECT. Springer, Cham (2021)
18. Kuzminykh. I., Carlsson A.: Analysis of Assets for Threat Risk Model in Avatar-Oriented IoT Architecture. NEW2AN/ruSMART 2018. LNCS, vol 11118, Springer, Cham (2018)
19. BenchmarkDotNet. Frequently asked questions. https://benchmarkdotnet.org/articles/faq.html. Last accessed 10 June 2020
20. Dooley, R., Edmonds, A., Hancock, DY., et al.: Security best practices for academic cloud service providers. Technical report (2018)
21. Topper, J.: Compliance is not security. Computer Fraud & Security, **2018**(3), 5–8 (2018)
22. Hashicorp. High Availability. https://www.vaultproject.io/docs/internals/high-availability.html. Last accessed 17 June 2020
23. Hashicorp. Production hardening. https://learn.hashicorp.com/vault/operations/production-hardening. Last accessed 17 June 2020
24. Openstack. Barbican Documentation. https://docs.openstack.org/barbican/latest/. Last accessed 17 June 2020
25. Pinterest. Knox Wiki. https://github.com/pinterest/knox/wiki/. Last accessed 17 June 2020
26. Square. Keywhiz. https://github.com/square/keywhiz. Last accessed 17 June 2020
27. Cyberark conjur automatically secures secrets used by privileged users and machine identities. https://github.com/cyberark/conjur. Last accessed 17 June 2020